# Global epistasis on fitness landscapes

Juan Diaz-Colunga[1]*, Abigail Skwara[1]*, Karna Gowda[2]*, Ramon Diaz-Uriarte[3]*, Mikhail Tikhonov[4]*, Djordje Bajic[1]* & Alvaro Sanchez[1,5]*#

[1] *Department of Ecology & Evolutionary Biology, Yale University, New Haven, CT, USA.*
[2] *Department of Ecology & Evolution & Center for the Physics of Evolving Systems, The University of Chicago, Chicago, IL, USA*
[3] *Department of Biochemistry, School of Medicine, Universidad Autónoma de Madrid, Madrid, Spain & Instituto de Investigaciones Biomédicas 'Alberto Sols' (UAM-CSIC), Madrid, Spain*
[4] *Department of Physics, Washington University of St. Louis, St. Louis, MO, USA*
[5] *Department of Microbial Biotechnology, Campus de Cantoblanco, CNB-CSIC, Madrid, Spain.*
* Equal contribution
# To whom correspondence should be addressed

JDC: juan.diazcolunga@yale.edu, ASk: abby.skwara@yale.edu, KG: karna.gowda@gmail.com, RDU: r.diaz@uam.es, MT: tikhonov@wustl.edu, DB: djordje.bajic@yale.edu, AS: alvaro.sanchez@cnb.csic.es

**Abstract:** Epistatic interactions between mutations add substantial complexity to adaptive landscapes, and are often thought of as detrimental to our ability to predict evolution. Yet, patterns of global epistasis, in which the fitness effect of a mutation is well-predicted by the fitness of its genetic background, may actually be of help in our efforts to reconstruct fitness landscapes and infer adaptive trajectories. Microscopic interactions between mutations, or inherent nonlinearities in the fitness landscape, may cause global epistasis patterns to emerge. In this brief review, we provide a succinct overview of recent work about global epistasis, with an emphasis on building intuition about why it is often observed. To this end, we reconcile simple geometric reasoning with recent mathematical analyses, using these to explain why different mutations in an empirical landscape may exhibit different global epistasis patterns — ranging from diminishing to increasing returns. Finally, we highlight open questions and research directions.

**Introduction**

Adaptive landscapes map genotypes to fitness under a given set of environmental conditions, and their topography has important consequences for the reproducibility and predictability of evolution. While originally formulated as a metaphor, the increase in experimental throughput over the past decades has made it empirically possible to quantitively map genotypes to phenotypes and fitness in a variety of systems. These include fitness landscapes in yeast and bacterial populations [1–3], the mapping between protein sequence and function [4–8], the mapping from RNA sequence to binding affinities for specific nucleotides [9,10], and the mapping from the sequence of transcription factor binding sites to their strength [11,12], among others. Besides their potential importance for predicting adaptation, knowing the topography of genotype-phenotype maps is essential if we wish to engineer biological function. In recent years, the idea of quantitatively mapping how biological function is encoded in the genetic composition of biological systems has been extended to those systems above the organismal level, such as ecological communities [13–18]. This work is part of an emerging paradigm in systems biology that seeks to generalize the landscape concept beyond genetics to describe and predict the function of combinatorially constructed biological systems, from antibiotic and cancer drug cocktails to stressors and metabolic networks [19–22].

The critical importance of landscape topography has prompted biologists to investigate how one may learn this topography from empirical data. The astronomical dimensionality of most combinatorial landscapes makes them highly challenging to map using empirical approaches alone. Computational tools exist to predict a range of molecular landscapes from first principles. The fold and thermal stability of small RNA molecules can be predicted from thermodynamic models [23], and the application of artificial intelligence to the longstanding problem of predicting protein structure from genotype has led to recent breakthroughs [24]. Despite this progress at the molecular level, our ability to predict biological function at the organismal level is more limited. Flux Balance Analysis (FBA) permits us to calculate various objective functions of metabolic networks, including the fitness of cells, from knowledge of the structure of these networks [25]. Models based on FBA have seen impressive success in predicting metabolic trait values to which adaptation would converge [26], and they have been shown to qualitatively capture the shape of metabolic fitness landscapes in model organisms such as *E. coli* [27–30]. Despite these success stories, the reach of FBA models is still limited, and many important organismal traits (from motility to cell shape) are outside of their scope. More sophisticated whole-cell models are being developed that may remedy this shortfall [31]. These promise to be highly valuable for the task at hand.

While prediction from microscopic models is challenging and not always possible, several methodologies have been proposed that seek to reconstruct the landscape from a subset of genotype-phenotype measurements using statistical inference tools [32–35]. These methods rely on the premise that regional and global features of landscapes are predictable from local information. If the association between genotype and phenotype were completely random (as in the house-of-cards landscape), we could not learn it from any subset of measurements, no matter how large, whereas if it were perfectly smooth and all mutations had additive effects (i.e., in a perfectly additive landscape), we would only need a small training subset that contains all the mutations we are interested in.

While, in principle, interactions between mutations (epistasis) might make landscapes so rugged that inferring the precise mapping between each genotype and its phenotype would become near impossible, several empirical studies have noticed that these interactions often follow systematic patterns that enable predictability [36,37]. Notably, epistasis (i.e., deviation from additive behavior) can have a global component, where the fitness (or phenotypic) effect of a mutation ($\Delta F$) is well-predicted by the fitness of the background where it is added ($F_B$). Patterns of global epistasis have been found in various fitness landscapes, including those of yeast [38–41] and bacterial [1,2,42] populations. More recently, we have identified that similar patterns are found in ecological landscapes, which map the genotypic composition of species assemblages to different ecological functions [18].

Perhaps the best-documented type of global epistasis is characterized by *diminishing returns*, where the fitness effect of a beneficial mutation becomes ever smaller in fitter backgrounds. This form of epistasis has been described in yeast [38,39,41] and bacterial populations [1,2,43,44], in multicellular organisms [45], as well as at the sub-cellular level, e.g., in protein landscapes [46,47]. It is also the most common form of global epistasis patterns in ecological community-function landscapes [18]. A second type of global epistasis, characterized by *increasing costs* has important implications for evolutionary robustness, and it describes the situation where the deleterious effect of a mutation increases in magnitude when added to fitter backgrounds [38]. While negative relationships between fitness effects and the fitness of the background have been more commonly reported, positive relationships have been found as well, where the fitness effect of a beneficial mutation increases in a predictable manner when added to fitter genetic backgrounds. This *increasing returns* epistasis has been reported in low-dimensional bacterial fitness landscapes [1,42], and it has also been described in systems at radically different scales of biological organization, from proteins [46] to ecosystems [18].

Beyond helping us predict the fitness or phenotypic effect of a mutation in a new background, the existence of global epistasis reflects regularities in the structure of genotype-phenotype maps, whereby "local" information on genotype-phenotype relationships may be extrapolated to predict global features of these relationships. These regularities are essential for predicting the topography of fitness landscapes, and they play a key role in several methodologies attempting to do this [32,35]. In what follows, we will provide an overview of our current understanding of global epistasis, with the hope of providing intuition about its origins.

**Global epistasis patterns are a function of the roughness of fitness landscapes.** In principle, global epistasis patterns might not seem inevitable. For example, in a purely additive landscape, where all mutations act independently, the fitness effect of a mutation ($\Delta F$) would be constant across all different genetic backgrounds regardless of their fitness ($F_B$), and there would be no relationship between the two (Fig. 1A). If interactions are present, it is not immediately clear if or how the fitness effect of a mutation will depend on the fitness of the genetic background. The relationship between $\Delta F$ and $F_B$ may simply be highly idiosyncratic, reflecting the particular genetic composition of each background. In this case, one might expect that the fitness effect of a mutation on backgrounds of varying fitness would be scattered more or less randomly around the average fitness effect of the mutation. As an extreme example, in the opposite limit to an additive, smooth landscape we have a maximally rugged landscape, where the association between genotype and fitness is essentially random. Should the fitness effect of a mutation correlate with the fitness of the background in this case? It indeed should, as a simple consequence of regression to the mean [48]. Any mutation that is added to the fittest genetic background will inevitably bring the fitness down (or, at best, it will keep it the same). By contrast, when we add that same mutation to the least-fit background in our landscape, fitness will have nowhere else to go but up. The same argument is true, statistically, when we add a mutation to genotypes of either higher or lower than average fitness, leading to a negative correlation between the fitness effect of every mutation and the fitness of the background to which we add them. As shown in Fig. 1B, all mutations in a random landscape should have the same, trivial scaling with slope $b = -1$ and coefficient of determination $R^2 \sim 0.5$. Therefore, in a random landscape, the fitness effect of a mutation always declines, on average, with the fitness of its genetic background. Although the effect is most extreme in random landscapes, regression to the mean will generally introduce a bias towards negative associations between $\Delta F$ and $F_B$ [49] (note that by a similar argument, biases can also be introduced by measurement errors; see [48]).

**Geometry gives us a simple interpretation of global epistasis patterns.** As discussed in the introduction, global epistasis can follow diminishing returns or increasing costs patterns, among others. One hypothesis is that these patterns emerge from interactions between mutations. How may these interactions give rise to either positive or negative associations between $\Delta F$ and $F_B$? To gain intuition about this question, it is instructive to examine the simplest possible situation: a close-to-additive landscape where most mutations act additively, and epistasis is as sparse as possible (Fig. 2A,B). The idea that sparse interactions can lead to the emergence of fitness-correlated trends, such as those typically connected to global epistasis, is considered and discussed in more detail in ref. [50]. Our goal here is to use this limit of sparse interactions as a vehicle for developing intuition, thanks to its simplicity.

Our starting point is the idealized additive landscape in Fig. 1A, composed of three beneficial and three detrimental mutations. In this otherwise additive landscape, we introduce, in Fig. 2, a single positive interaction between two of the beneficial mutations, *mut. 1* and *mut. 2,* so that $\varepsilon_{12} > 0$ as depicted in the upper right quadrant of Fig. 2B. This interaction will create two distinct types of backgrounds to which *mut. 1* can be added: Those that contain *mut. 2*, and those that do not. Because *mut. 2* is beneficial and purely additive in the absence of *mut. 1*, backgrounds that contain *mut. 2* (orange in Fig. 2C) will tend to have higher fitness, on average, that those that do not (gray in Fig. 2C). This leads to a horizontal shift between gray (*mut. 2*-) and orange (*mut. 2*+) backgrounds, whose magnitude is given by the fitness effect of *mut. 2* in the genetic backgrounds lacking *mut. 1* ($\delta_2$). Because epistasis is positive, the fitness effect of adding *mut. 1* to the orange (*mut. 2*+) backgrounds will be larger than when we add it to gray backgrounds (lacking *mut. 2*) by an amount equal to the epistatic interaction ($\varepsilon_{12}$). There is, therefore, an upwards vertical shift between gray (*mut. 2*-) and orange (*mut. 2*+) backgrounds as a result of adding *mut. 1* to both. The combination of a rightwards shift, towards genetic backgrounds of higher fitness when *mut. 2* is present (caused by its positive fitness effect), and an upwards shift when *mut. 2* is present (caused by positive epistasis between *mut. 2* and *mut. 1*) will produce a positive regression slope between $\Delta F$ and $F_B$ which magnitude will be proportional to $\varepsilon_{12}/\delta_2$ (Fig. 2C, upper right quadrant). Now, using a similar argument, we can see that if the interaction between *mut. 1* and *mut. 2* were instead negative (i.e., if $\varepsilon_{12} < 0$), the vertical shift in $\Delta F$ between gray and orange dots would be downwards (Fig. 2C, lower right quadrant). The combination of a downwards vertical shift and a rightwards horizontal shift in the orange (*mut. 2*+) dots will lead to a negative slope (Fig. 2C, lower right quadrant). Importantly, we note that the additive fitness effect of *mut. 1* does not matter as far as the slope is concerned. This is because it will apply equally to all backgrounds when we introduce *mut. 1* and, therefore, it will not cause a vertical shift in just some of the backgrounds but not others (i.e., a different $\delta_1$ would result in the exact same figure, but with all points shifted left or right).

The same geometric logic can be applied to explain the emergence of linear relationships between $\Delta F$ and $F_B$ as a result of either negative (antagonistic) or positive (synergistic) pairwise epistasis between mutations of opposite fitness effects (e.g., one positive and one negative, as shown in Fig. 2B-C in the remaining quadrants). In sum, in the simplest possible landscapes where a single pairwise interaction exists between two mutations, a positive slope between $\Delta F$ and $F_B$ should be seen when the signs of epistasis and fitness effects of the background mutation are aligned (i.e., both positive or both negative). By contrast, when these signs are not aligned, a negative slope would arise.

The simple fitness landscape discussed above also emphasizes that global epistasis may arise statistically as we plot together different "regional" trends that capture the relationship between $\Delta F$ and $F_B$ in different regions of the genotype space. The possibility that sparse interactions can give rise to the emergence of a fitness-correlated trend has been empirically examined in recent work [50]. By studying a complete fitness landscape made up by ten different mutations assessed in six different environments, the authors found several mutations in yeast that were consistent with this pattern, concluding that a small number of low-order epistatic interactions are sufficient to cause global epistasis-like trends.

**Global epistasis may also arise from widespread microscopic interactions between mutations.** The observations stemming from our analysis of the close-to-additive landscape (Fig. 2) are overall consistent with recent analyses by Lyons et al. [40] and Reddy & Desai [49]. Both works showed how diminishing returns and increasing costs epistasis can arise as a generic consequence of widespread (rather than sparse) microscopic interactions. Because the presence of many high-order interactions can cause the effect of mutations to depend sensitively on the genetic background, a high degree of this so-called "idiosyncratic epistasis" means that the fitness of neighboring genotypes are essentially uncorrelated (as in Fig. 1B). This implies, as a consequence of regression to the mean, that the effects of mutations to high-fitness genotypes are more likely to be small if positive (diminishing returns) and large if negative (increasing costs). Lyons et al. [40] demonstrated that, across systems, idiosyncratic epistasis is common. The authors defined an empirical "idiosyncratic index" for fitness landscapes, which captures the variability of mutational effects to different genetic backgrounds. In a number of fitness landscapes, ranging from proteins to organisms, the authors found evidence for a moderate to high degree of idiosyncratic epistasis, suggesting that widespread microscopic interactions may underlie patterns of global epistasis in many different systems.

This recent work also suggests that the intuition gained in Fig. 2 may still be useful in complex landscapes, where microscopic epistasis is not sparse. As shown in Box 1, it follows from the mathematical analyses in Reddy & Desai [49] that the slope of the linear regression between $\Delta F$ and $F_B$ for a given mutation (say *mut. 1*) is given by the sum over all other mutations ($j \neq 1$) of the product between $\langle \delta F_j \rangle_{B(1)}$ (the average fitness effect of mutation *j* over all backgrounds on which we may place *mut. 1*, denoted B(1)), and $\langle \varepsilon_{1j} \rangle$ (the average epistasis between mutation 1 and mutation *j* over all possible backgrounds where we may add this pair). As explained in Box 1, $\langle \delta F_j \rangle_{B(1)}$ quantifies the average horizontal shift in $F_B$ caused by adding mutation *j* to the backgrounds of mutation 1, like $\delta_2$ did in Fig. 2. Meanwhile, $\langle \varepsilon_{1j} \rangle$ characterizes the average vertical shift in ΔF when we add *mut. 1* to those very same backgrounds. Consistent with the geometric intuition given in Fig. 2, when the signs of both of these terms are aligned for a particular mutation *j*, this mutation will contribute to increasing the slope, making it more positive. By contrast, when their signs are not aligned, mutation *j* will contribute to a negative slope. The sign of the final slope of the global epistasis pattern for our focal mutation *mut. 1* will be determined by the sign of the sum of $\langle \varepsilon_{1j} \rangle \times \langle \delta F_j \rangle_{B(1)}$ over all other mutations *j* with which *mut. 1* may interact.

**Empirical support for the microscopic epistasis model.** The analysis by Reddy & Desai [49], which we briefly summarize above, quantitatively connects the slope of global epistasis to the set of all microscopic interactions between mutations. Moreover, it makes precise predictions about the distribution of residuals around the global epistasis patterns. The authors find good agreement between these predictions and the results of an earlier experiment in the same lab, where global epistasis was measured for 91 mutations introduced in a library of different yeast genetic backgrounds [38]. This suggests that the model assumptions (namely, a large number of independent interactions between loci and additive effects that are smaller than epistatic interactions) may be fulfilled for this particular system.

While the validity of these assumptions is important for modeling the distribution of residuals, the limit of a large number of independent interactions is not required for the theory to explain the sign of global epistasis. As an example, in Fig. 3 we show the expected global epistasis patterns in the fitness landscape studied by Khan et al., which includes just five beneficial mutations from a long-term adaptation experiment in *E. coli* [1]. The model in ref. [49], truncated up to pairwise epistasis (reproduced in Box 1) does an excellent job at explaining how the global epistasis pattern for each mutation depends on the microscopic interactions between mutations (Fig. 3). Because this fitness landscape is complete, and includes mutations that exhibit positive, negative, and near zero global epistasis slopes, we can empirically determine what kinds of interactions and fitness effects give rise to the sign of each slope.

To this end, we show in Fig. 3 a breakdown of every component that contributes to the slope of global epistasis for each mutation. That is, for each focal mutation, we show the epistatic interactions with each other mutation (averaged over every possible background) as well as the average fitness effect of each other mutation on the genetic backgrounds of our focal mutation. The picture that emerges is consistent with the geometric reasoning in Fig. 2. For instance, the mutation involving the pyruvate kinase gene *pykF* exhibits a clear pattern of increasing returns epistasis (Fig. 3B, rightmost panel). This mutation, as all others in this experimental landscape, is beneficial and therefore it has a positive effect, on average, in every background [1]. Based on the geometric reasoning above, we would expect this behavior to emerge from positive pairwise interactions with the other mutations, such that the pairwise epistasis is aligned in sign with the fitness effects. This is indeed the case: all interactions involving *pykF* are indeed synergistic and epistasis is positive, as already noted in the original study [1]. By contrast, the three mutations that exhibit diminishing returns epistasis (those in the *topA* and *spoT* genes, and in the *glmUS* promoter) have a dominant negative epistatic interaction with the mutation that has the strongest fitness effect in their respective backgrounds. This misalignment between the signs of epistasis (negative) and the fitness effect of the dominant background interaction (positive) explains the negatively sloped regression between ΔF and $F_B$ in these three genes. Finally, the mutation on the *rbs* operon exhibits one negative interaction with the one other mutant that causes the strongest horizontal shift in the fitness of its genetic background (*glmUS*), and three milder positive interactions with the rest of the mutations. These partially compensate for the negative contribution of *glmUS*, leading to a slope that is close to zero.

Through this example, we intend to illustrate that the geometric reasoning in Fig. 2 and the mathematical analyses in ref. [49] can offer valuable intuition about the microscopic origins of the global epistasis patterns observed in empirical fitness landscapes. While we have focused on the slope, the intercept can be analyzed in a similar manner and can be explained from microscopic interactions as well (see Box I).

**Global epistasis as a non-linear transformation of a latent additive variable.** An alternative view of global epistasis that does not invoke microscopic epistasis is that it may arise from a non-linear transformation of a latent, additive variable to which mutations contribute [12,51]. Nonlinear transformations of an additive variable may indeed give rise to higher-order epistasis [52], and their contribution to global epistasis has been explored in a range of different landscapes, including protein [46] and antibody-antigen affinity landscapes[53]. Learning these nonlinear transformations from data underlies recent methodologies for inferring fitness landscapes from a subset of measured phenotype-genotype pairs [32].

One may readily see that, indeed, non-linear transformations on an additive landscape may produce correlations between the fitness effect of a mutation and the fitness of the background. As a simple exercise, in Fig. 4, we apply either a convex or a concave exponential transformation ($F=\exp(\lambda)$ or $F=1-\exp(-\lambda)$, respectively) to an idealized, additive mapping between genotypes and a latent additive variable ($\lambda$), which may be also referred to as the fitness potential [54,55]. After the transformation, patterns consistent with the definition of global epistasis given above readily emerge. As one would have expected, a convex exponential transformation leads to increasing returns for beneficial mutations, and to increasing costs for detrimental mutations. Both of these go together, reflecting the low robustness of the resulting landscape [38]. By contrast, the concave transformation leads to diminishing returns and decreasing costs, reflecting the increased robustness of the underlying landscape (we do not mean to imply, however, that either scenario reflects a natural landscape).

It is easy to show mathematically (and illustrated graphically in Fig. 4) that exponential transformations generate linear dependencies between $\Delta F$ and $F_B$. This is however not true for all transformations, though measurement error and other sources of stochasticity might hide the non-linearity of the $\Delta F$ vs. $F_B$ relationship. Although non-linear transformations can lead to global epistasis patterns, such as diminishing and increasing returns (Fig. 4) they produce distinct patterns of residuals from those observed when the source of global epistasis are idiosyncratic microscopic interactions between mutations, making it possible to distinguish between the two mechanisms empirically [49]. Most notably, epistatic interactions become correlated, breaking the assumption of independence that underlies the model in ref. [49]. In addition, the shape of the nonlinear transformation may put constraints on which form of microscopic epistasis may be observed. For instance, monotonic transformations are not compatible with sign epistasis, which would require non-monotonic (e.g., unimodal) transformations with at least one peak at an intermediate value of the parameter $\lambda$. Alternatively, sign epistasis may arise when a non-linear transformation is applied to a vector of additive variables, rather than to a scalar, additive parameter $\lambda$. This may be leveraged to help us discriminate between global epistasis arising from a nonlinear transformation versus arising from microscopic interactions between mutations.

The view of global epistasis as emerging from a nonlinear transformation of a latent additive variable has received additional theoretical support from recent work by Husain & Murugan [36]. This work reports that the biophysical dynamics of biological systems may constrain epistasis, in systems ranging from protein landscapes to gene regulatory networks. For this to occur, a slow dynamic mode must exist that dominates the dynamics of the system at long timescales, when all of the faster modes have decayed already. Under this regime, the mutational and epistatic effects of different mutations lose statistical independence, leading to global epistatic patterns.

Despite their differences, both views of global epistasis have been found to fit and predict empirical features in different data sets [32,46,53]. Fully understanding the scope of both models and identifying the situations when either of the two would be most appropriate is an open area of inquiry that will require additional work.

**Open questions:** We have discussed global epistasis among mutations, but what do we mean exactly by a mutation? In some cases, a mutation may be thought of as a simple substitution, while in others it may refer to the deletion of a whole gene. Do global epistasis patterns change substantially at different scales of coarse-graining? For instance, plasmids can be transformed into different genetic backgrounds, and their fitness effects may then be determined in each [56]. Do global epistasis patterns extend to the gain of thousands of kb of plasmid DNA via horizontal gene transfer? If so, which patterns are the most common, and how do these influence the propagation of mobile elements? It remains to be determined whether the patterns of global epistasis change as we zoom in and out the scale of what we identify as a mutation, from the base pair level to the gene level, and from the gene level to the operon and beyond.

A related question is the prevalence of regional epistasis. The idea that the relationship between the fitness effect of a mutation and the fitness of the background may be different in different regions of the landscape is hinted at in recent empirical data, both in genetic landscapes (e.g., [38,50]) and in ecological community-function landscapes [18]. We are just beginning to understand how this regional epistasis is connected to global patterns. It is possible, for instance, that spurious global patterns can be generated by undersampling a landscape with regional epistasis patterns [50]. We are just beginning to understand how this regional epistasis is connected to global patterns. It is possible, for instance, that global patterns are seen by undersampling regional epistasis patterns. How one may learn these regional trends from data remains to be determined. Likewise, the question of whether the size of the landscape matters for the kind of global epistasis observed is an important one that is not yet fully settled. Connecting these regional patterns with theory and mechanism is in its early days, and this is an area where a combination of theory and experiments is likely going to be needed.

A third important question is the relative importance of nonlinear transformations on latent additive variables to global epistasis. It is likely that the answer will be different in different systems and for different phenotypes. Under which conditions should one expect the effect of nonlinear transformations to dominate global epistasis patterns, and when should we expect to see these patterns arising primarily as a result of microscopic interactions? Both models are not exclusive and in fact both may coexist to different degrees in different systems [53]. Whatever the answer to this question is, an immediate follow up direction is to determine the biophysical mechanisms that may be responsible. For instance, if one (or more) [54] latent variable is found to exist that maps non-linearly to the phenotype or fitness under study, we should want to determine what magnitude it represents and how the observed nonlinear mapping emerges from biophysical and biochemical principles. In other words, we would need to find out what is the latent variable (or variables) and why does the nonlinear transformation have the shape that it does. Computational models may give us some insight into this question in particular systems and scales of landscape description, even if a full systematic answer may not exist.

Finally, the recent observation of global epistasis-like patterns in ecological community-function landscapes suggests that these may be observed in other biological systems as well. Encouragingly, mathematical analyses such as those in ref. [49] are very general and do not necessarily assume that the interacting elements must be genes. Therefore, we hypothesize that global epistatic patterns may be leveraged to learn the mapping between composition and function of combinatorial landscapes of drugs and antibiotics, stressors, and, more generally, the collective performance of biological systems at different levels of organization.

**Conclusion:** Because our goal was primarily to provide a brief summary and intuition, our overview has not been comprehensive, and we have not been able to properly discuss many outstanding issues in the study of epistasis. Besides the open questions raised above, there exist many other aspects that deserve further investigation. For instance, some of the evidence for global epistasis (such as that presented in Fig. 3) comes from small combinatorial landscapes. The choice of which mutations are included in these finite landscapes can have important consequences for the type of epistatic patterns observed. As convincingly argued elsewhere [51,57], if the choice of mutations includes only beneficial mutations that fixed sequentially to form a highly fit genotype due to natural selection, we will be basing our conclusions in that particular slice of the fitness landscape. This slice contains a highly non-random set of mutations and therefore is unlikely to be representative of patterns observed more generally. Indeed, theoretical arguments have been made that epistasis may be weaker far away from highly fit genotypes, and become stronger and more often positive near a fitness peak [51,57,58]. Thus, the observed patterns of global epistasis can be biased by which mutations we include in the landscape.

How exactly this choice impacts the global epistasis patterns observed is an important quantitative question. For instance, negative epistasis appears to be more common in empirical landscapes than positive epistasis (e.g., [2,5,59]), and the most frequent global epistatic pattern is diminishing returns (e.g., three out of five mutations in ref. [1], four out of five mutations in ref. [2]). Incidentally, the same is true in ecological landscapes, where only a minority of species exhibiting global epistasis follow a pattern other than diminishing returns [18]. However, the theoretical work from ref. [49] (Box 1), indicates that one would have seen a higher prevalence of increasing returns global epistasis patterns if the chosen mutations were enriched in positive epistasis. As positive epistasis appears to be more common near fitness peaks [58], the choice of fitter genetic backgrounds may also matter in this regard. The availability of new experimental tools for the high-throughput characterization of fitness landscapes (e.g., [50,55]) and recent theoretical developments [49] can provide additional insights into this matter.

We still have much to learn about the phenomenon that has been termed global epistasis, but the payoffs of a more comprehensive understanding would be large. At the conceptual level, such an understanding will bring us closer to learning how genetic interactions affect the reproducibility and predictability of adaptation [39,40,59]. Global epistasis has also been instrumental in the development of methodologies that infer fitness landscapes from an incomplete set of observations [32,53]. Recently, this very idea has been extended beyond genetics to predicting ecological community-function landscapes from a subset of assembled communities [18]. In sum, methods exploiting global epistasis may be leveraged to engineer the function of biological systems at scales ranging from molecules to ecosystems, and they may also guide our efforts in the important global challenge of forecasting adaptive evolution [60,61]. Future work on this topic is thus likely to have wide implications across many fields of research.

**Data Availability**

The code used to generate all figures can be accessed at [github.com/jdiazc9/global_epistasis_review](github.com/jdiazc9/global_epistasis_review).

**Acknowledgements**

RDU was supported by grant PID2019-111256RB-I00 funded by MCIN/AEI/10.13039/501100011033.

## Glossary

**Global epistasis:** Describes a situation when the fitness effect of mutations depends on the other mutations that are present, "but only through their combined effect on fitness" [39].

**Diminishing returns epistasis:** Beneficial mutations have weaker fitness effects in fitter genetic backgrounds.

**Diminishing costs epistasis:** Detrimental mutations have weaker fitness effects when placed in fitter genetic backgrounds.

**Increasing returns epistasis:** Beneficial mutations have stronger fitness effects in fitter genetic backgrounds.

**Increasing costs epistasis:** Detrimental mutations have stronger fitness effects in fitter genetic backgrounds.

**Positive epistasis:** The situation where the effect of N>1 mutations is larger than expected if their effects on the observed phenotype or fitness were independent.

**Negative epistasis:** The situation where the effect of N>1 mutations is smaller than expected if their effects on the observed phenotype or fitness were independent.

**Modular epistasis model:** mutations affecting the same pathway may be functionally redundant, leading to diminishing effects in combination.

**Box 1**

As shown in ref. [49] the fitness of a genotype can be written as:

$$F = \overline{F} + \sum_i f_i x_i + \sum_{j>i} f_{ij} x_i x_j + \sum_{k>j>i} f_{ijk} x_i x_j x_k + \cdots$$

where $x_i = \pm 1$ reflects the presence or absence of mutation $i$, and the parameters $f_i, f_{ij}, f_{ijk}$ represent additive fitness contributions and the pairwise and higher order interactions.

From this model, and by neglecting interactions of order higher than two, the slope of the regression $b_i$ between the fitness effect of mutation $i$ ($\Delta F_i$) and the fitness of its genetic background ($F_B$) can be approximated as [49]:

$$b_i \approx 2 \frac{\sum_{j \neq i} f_{ij}(f_j - f_{ij})}{\sum_{j \neq i}(f_j - f_{ij})^2},$$

whereas the intercept of the regression takes the form $a_i = \langle \Delta F \rangle - b_i \langle F_B \rangle$. According to the equation above, the sign of the slope $b_i$ will be thus given by the sign of $\sum_{j \neq i} f_{ij}(f_j - f_{ij})$.

The slope of the regression can also be written in terms of the fitness values of every genotype $F(x_1, x_2, ...)$ instead of the interaction parameters ($f_i, f_{ij}, f_{ijk}, ...$). After some basic algebraic manipulations, and using the fact that: $f_{ij} = \langle \varepsilon_{ij} \rangle / 4$ and $(f_j - f_{ij}) = \langle \delta F_j \rangle_{B(i)}/2$ [49], one finds that the slope of the regression between $\Delta F_i$ and $F_B$ can be written as:

$$b_i \approx \frac{\sum_{j \neq i} \langle \varepsilon_{ij} \rangle \times \langle \delta F_j \rangle_{B(i)}}{\sum_{j \neq i} \left(\langle \delta F_j \rangle_{B(i)}\right)^2}$$

where $\langle \varepsilon_{ij} \rangle$ denotes the average pairwise epistasis between mutations $i$ and $j$ over every background where this pair may be added (e.g., for a landscape containing just two mutations, 1 and 2, $\varepsilon_{12} = F_{12} - F_1 - F_2 + F_0$), and $\langle \delta F_j \rangle_{B(i)}$ denotes the average fitness effect of mutation $j$ over the set of all genetic backgrounds where the focal mutation $i$ may be added (which we have denoted as $B(i)$).

The equation above can be written as

$$b_i \approx \sum_{j \neq i} \omega_{ij} \times \beta_{ij}$$

where we have defined

$$\omega_{ij} \equiv \frac{\left(\langle \delta F_j \rangle_{B(i)}\right)^2}{\sum_{j \neq i} \left(\langle \delta F_j \rangle_{B(i)}\right)^2}$$

and

$$\beta_{ij} \equiv \frac{\langle \varepsilon_{ij} \rangle}{\langle \delta F_j \rangle_{B(i)}}$$

Note that $\beta_{ij}$ can be interpreted as a generalized form of the expression for the slope in the simple landscape discussed in Fig. 2 (where the slope was given by $(\varepsilon_{ij}/\delta_j) \times v_j$, and $v_j$ is the fraction of the total variation in $F_B$ that is caused by the fitness effect of mutation $j$). The coefficient $\beta_{ij}$ may be seen as the individual contribution of mutation $j$ to the slope of the global epistasis pattern of mutation $i$. Said slope thus results from a sum of contributions from every mutation $j \neq i$, each weighted by a factor $\omega_{ij}$ which captures the relative magnitude of the fitness effect of mutation $j$ on the backgrounds of mutation $i$.

Up to second order epistasis, the sign of the slope is given by the sign of the sum: $\sum_{j \neq i} \langle \varepsilon_{ij} \rangle \times \langle \delta F_j \rangle_{B(i)}$. It is clear then that the sign of the slope for a focal mutation has contributions from its interactions with every other mutation $j$. Importantly, every such contribution is given by the product $\langle \varepsilon_{ij} \rangle \times \langle \delta F_j \rangle_{B(i)}$. This reveals a simple geometric interpretation that connects with that given in Fig. 2. The term $\langle \delta F_j \rangle_{B(i)}$ represents the average horizontal shift in the fitness of the background genotypes $F_B$ caused by mutation $j$, whereas the term $\langle \varepsilon_{ij} \rangle$ represents the average vertical shift in $\Delta F_i$ caused by interactions between $i$ and $j$ in different genetic backgrounds. Consistent with the geometric argument in Fig. 2, any mutation $j$ for which both $\langle \varepsilon_{ij} \rangle$ and $\langle \delta F_j \rangle_{B(i)}$ have the same sign will contribute towards a positive global epistasis slope, whereas those for which the two terms have different signs will contribute to making the global epistasis slope more negative. This highlights that both the average fitness effect and the average epistatic effect matter equally. A mutation with a very strong fitness effect over the backgrounds of the focal mutation may dominate the sign of the slope even if its epistatic interactions with the focal mutation are not the strongest. In sum, the mathematical analysis in ref. [49] generalizes and formalizes the simple geometric intuition given in Fig. 2.

**Figures**

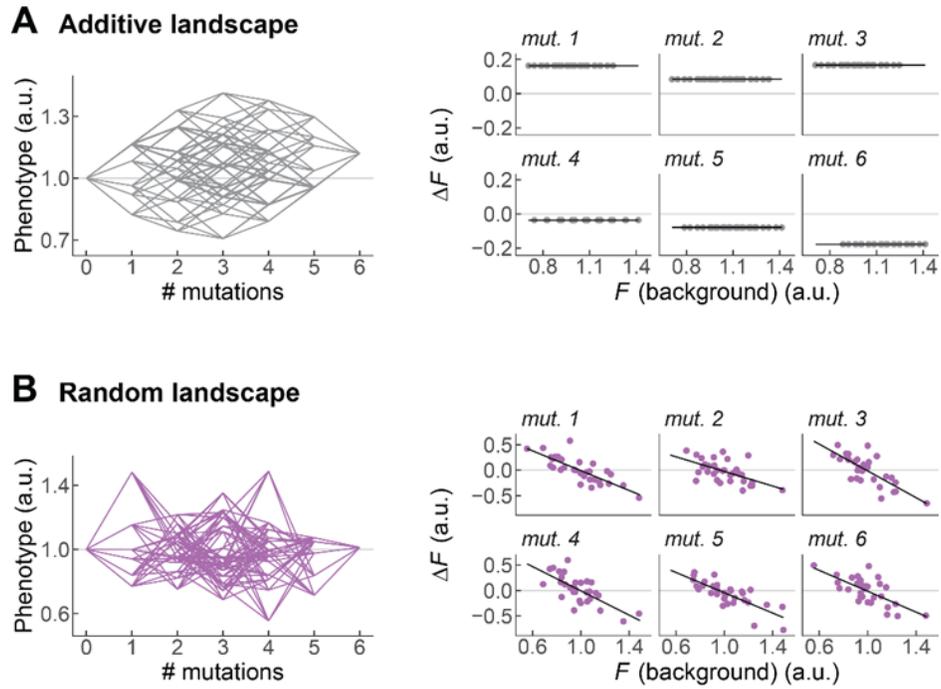

**Fig. 1. The ruggedness of a fitness landscape determines the shape of global epistasis patterns.** We represent two limit cases for the topography of a fitness landscape of six mutations. In the right panels, we represent the fitness effect of each mutation ($\Delta F$) as a function of the fitness of the genetic background where it is introduced ($F_B$). The fitness effect of mutation $i$ in the background $B$ is defined as $F_{B+i} - F_B$, where $B+i$ represents the genotype resulting from the addition of mutation $i$ to the genetic background $B$. (**A**) A maximally smooth landscape where all mutations act additively produces flat global epistasis patterns. (**B**) A rugged landscape in which every genotype is sampled independently from an arbitrary distribution (in this example, a normal distribution with mean 1 and standard deviation 0.2) results in global epistasis patterns in which every mutation displays a slope around -1 and an intercept around the average fitness of the landscape. Black lines represent least-square fits. a.u.: arbitrary units.

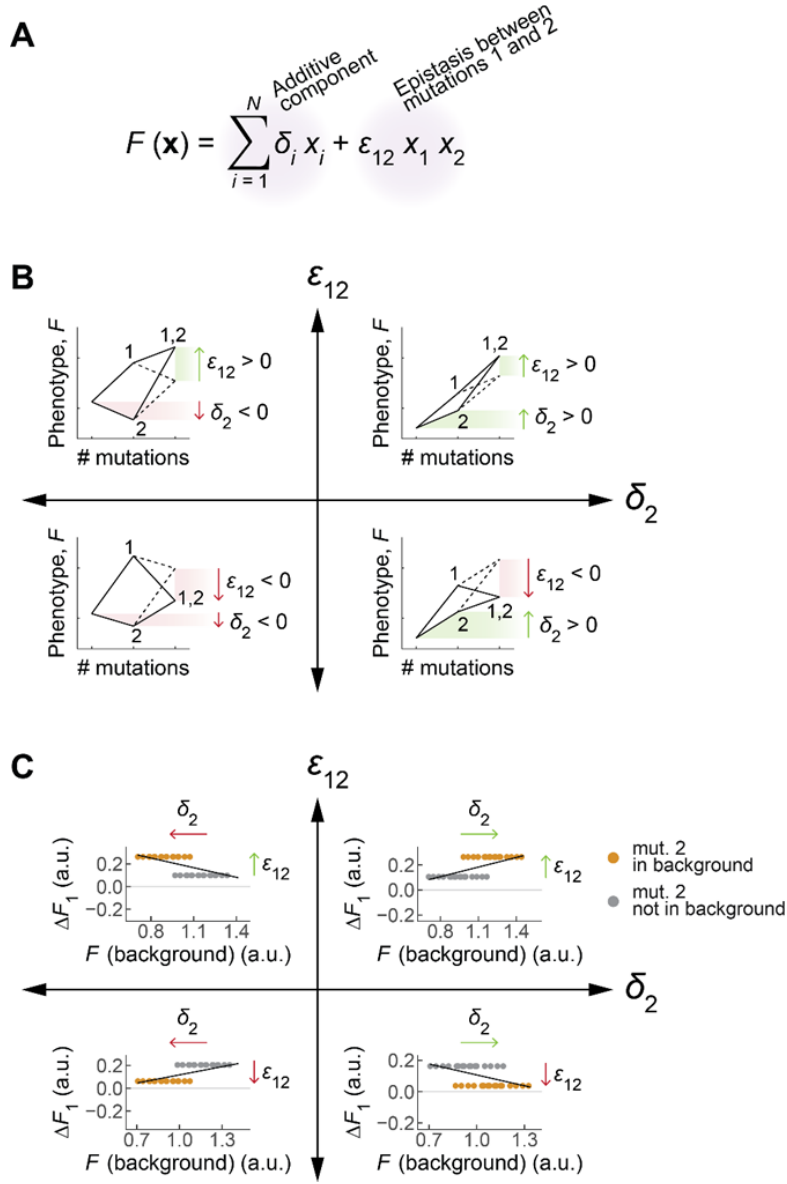

**Fig. 2. Global epistasis in a simple landscape with a single pairwise interaction.** (**A**) We consider a simple model in which the fitness (or phenotype) $F$ of an organism is given by the sum of an additive contribution from every mutation ($\delta_i$ for mutation $i$, here $N=6$ mutations in total) and an epistatic term capturing a pairwise interaction between mutations 1 and 2 ($\varepsilon_{12}$). In this formulation, $x_i$ is either 1 (if mutation $i$ is present) or 0 (if it is absent). (**B**) The sign of $\delta_2$ determines whether, in the absence of mutation 1, mutation 2 has a beneficial or deleterious effect. The sign of $\varepsilon_{12}$ determines the direction of fitness change when both mutations 1 and 2 are present with respect to the additive case. (**C**) The signs of $\delta_2$ and $\varepsilon_{12}$ dictate the slope of the trend observed between the fitness effect of mutation 1 ($\Delta F_1$) and the fitness of the genetic background. Note that the sign of $\delta_1$ does not affect the sign of the slope for the global epistasis pattern of mutation 1 itself (in the examples shown, $\delta_1 > 0$, so acquiring 1 in the absence of 2 leads to an increase in $F$). The reason is that the slope is caused by a vertical shift in the plots above, which in turn is caused by the fitness effect of mutation 1 ($\Delta F_1$) being different in backgrounds that also include mutation 2 (*mut.2+* backgrounds) or those that lack it (*mut.2-* backgrounds). In this simple scenario, $\Delta F_1$ in *mut.2+* backgrounds is given by $\Delta F_1(\textit{mut.2+}) = \delta_1 + \varepsilon_{12}$ whereas in *mut.2-* backgrounds we have simply $\Delta F_1(\textit{mut.2-}) = \delta_1$. The vertical shift between gray and orange dots is given by $\Delta F_1(\textit{mut.2+}) - \Delta F_1(\textit{mut.2-}) = \varepsilon_{12}$. Therefore, neither the sign nor magnitude of $\delta_1$ have any impact on the regression slope between $\Delta F_1$ and $F_B$. The choice of a positive $\delta_1$ in this panel is thus arbitrary and has no impact on the results.

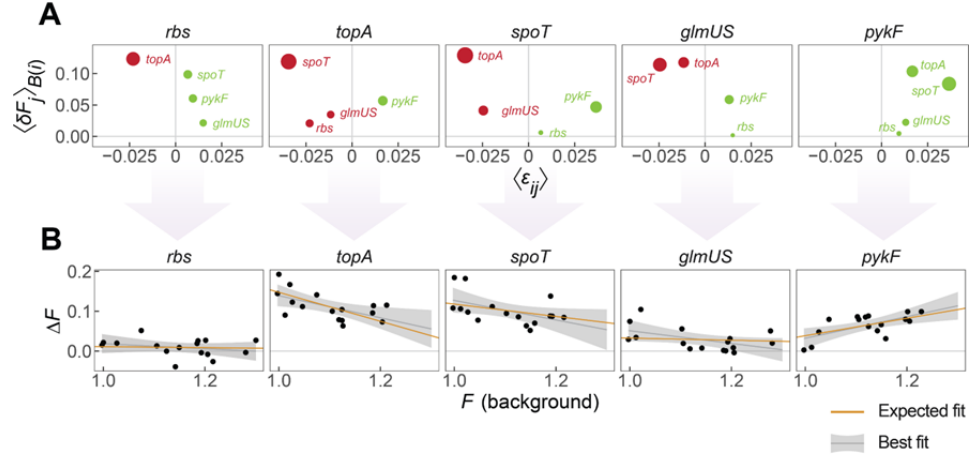

**Fig. 3. Microscopic interactions can explain the slope of global epistasis.** We consider a landscape of five beneficial mutations from the *E. coli* long-term evolution experiment [1]. (**A**) Each panel corresponds to a different mutation (the *focal*). Each dot corresponds to a pair of said focal mutation with each of the other four mutations in the data set. In the vertical axis, we represent the average fitness effect of the non-focal mutations when they are added to the set of genetic backgrounds that do not contain the focal mutation — we denote this set of backgrounds as $B(i)$, where $i$ is the focal mutation. Note that this average, $\langle \delta F_j \rangle_{B(i)}$, becomes just the coefficient $\delta_2$ in the simpler scenario of Fig. 2. In the horizontal axis, we represent the average pairwise epistasis between the focal mutation and the indicated non-focal mutation, $\langle \varepsilon_{ij} \rangle$, across every possible background that does not contain either of them (note the analogy with $\varepsilon_{12}$ in Fig. 2). Dots colored in green/red correspond to instances where this average pairwise epistasis is positive/negative. The size of the dots is proportional to the product of $\langle \varepsilon_{ij} \rangle \times \langle \delta F_j \rangle_{B(i)}$ (see Box 1). (**B**) The slopes and intercepts of the global epistasis patterns for each focal mutation can be estimated from a sum of such $\langle \varepsilon_{ij} \rangle \times \langle \delta F_j \rangle_{B(i)}$ terms as described in Box 1. Note that, in this data set, all five mutations are beneficial ($\langle \delta F_j \rangle_{B(i)} > 0$ for every pair of mutations $i$ and $j$). When negative pairwise epistasis dominates, the slopes are negative (e.g., mutations *topA* or *spoT*). When pairwise epistasis is predominantly positive, the slope is positive (mutation *pykF*). When positive and negative terms have roughly the same weight, the slope is flat (mutation *rbs*). The equations in Box 1 can provide quantitative estimates for the slopes (yellow lines), which agree with the least-squares fits to the data (gray lines, shadowed regions represent 95% confidence intervals around the least-squares fit). This reflects that, at least in this system, the assumption of neglecting higher-than-pairwise terms is a good approximation.

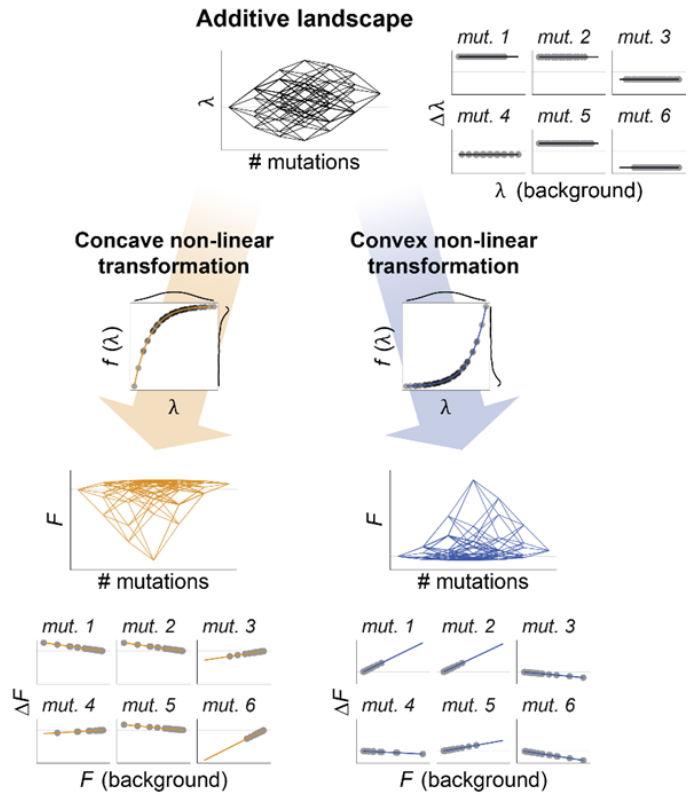

**Fig. 4. Global epistasis patterns can emerge from a non-linear transformation on an additive landscape.** Here, we represent an example of a landscape for a latent additive variable that we denote as $\lambda$. We consider two cases where the fitness $F$ is a non-linear function of $\lambda$, namely $f(\lambda)$. In the first case, $f$ is a concave function of the form $1 - \exp(-\lambda)$ (orange); in the second, it is a convex function of the form $\exp(\lambda)$ (blue). The concave transformation results in the emergence of *diminishing returns* patterns (a form of negative epistasis) for the mutations with beneficial effects on $\lambda$ (mut. 1, mut. 2 and mut. 5), and *decreasing costs* patterns (a form of positive epistasis) for those mutations with deleterious effects on $\lambda$ (mut. 3, mut. 4, and mut. 6). The convex transformation leads to patterns of *accelerating returns* (another form of positive epistasis) for the mutations with beneficial effects on $\lambda$ and to *increasing costs* (a form of negative epistasis) for those mutations with deleterious effects on $\lambda$.